\documentclass[3p,times,twocolumn]{elsarticle}

\usepackage{ecrc}

\newcommand{\be}{\begin{equation}}
\newcommand{\ee}{\end{equation}}

\volume{00}

\firstpage{1}

\journalname{Nuclear Physics B Proceedings Supplement}

\runauth{}

\jid{nuphbp}

\jnltitlelogo{Nuclear Physics B Proceedings Supplement}

\usepackage{amssymb}

\usepackage[figuresright]{rotating}

\begin{document}

\begin{frontmatter}



\dochead{}

\title{Testing the Zee-Babu model via neutrino data, lepton flavour violation
and direct searches at the LHC}


\author{Juan Herrero-Garcia$^{a}$, Miguel Nebot$^b$, Nuria Rius$^c$ and Arcadi Santamaria$^c$}

\address{$^a$ Department of Theoretical Physics,
School of Engineering Sciences, KTH Royal Institute of Technology,\\
AlbaNova University Center, Roslagstullsbacken 21, 106 91 Stockholm, Sweden\\

$^b$ Centro de F\'isica Te\'orica de Part\'iculas, Instituto Superior T\'ecnico -- Universidade de Lisboa,\\ Av. Rovisco Pais 1, 1049-001 Lisboa, Portugal\\

$^c$ Departamento de F\'isica Te\'orica, Universidad de Valencia and IFIC, Universidad de Valencia-CSIC,\\
C/ Catedr\'atico Jos\'e Beltr\'an, 2 | E-46980 Paterna, Spain}

\begin{abstract}
In this talk we discuss how the Zee-Babu model can be tested combining information from neutrino data, low-energy experiments and direct searches at the LHC. We update previous analysis in the light of the recent measurement of the neutrino mixing angle 
$\theta_{13}$ 
\cite{An:2012eh}, the new MEG limits 
on $\mu\rightarrow e \gamma$ \cite{Adam:2013mnn}, the lower bounds on doubly-charged scalars coming from LHC data \cite{ATLAS:2012hi,Chatrchyan:2012ya},
and, of course, the discovery of a 125 GeV Higgs boson by ATLAS and CMS \cite{Aad:2012tfa,Chatrchyan:2012ufa}. In particular, we find that the new singly- and doubly-charged scalars are accessible at the second run of the LHC, yielding different signatures depending on the neutrino hierarchy and on the values of the phases. We also discuss in detail the stability of the potential.
\end{abstract}

\begin{keyword}
Neutrino masses \sep Lepton flavor violation \sep LHC \sep stability of the potential

\end{keyword}

\end{frontmatter}


\section{Introduction} \label{}

Radiative models are a very plausible way in which neutrinos may acquire their tiny masses: $\nu$'s are light because they are massless at tree level, with their masses being generated by loop corrections  that generically have the following form:
\be
 m_\nu \sim c\,\frac{v^2}{(4\pi)^{2i}\Lambda},
\ee
where $c$ encodes some lepton number violating (LNV) couplings and/or ratios of masses, $\Lambda$ is the scale of LNV which can be at the TeV and therefore can be accessible at colliders, and $i$ are the number of loops, where typically more than three loops yield too light neutrino masses or have problems with low-energy constraints (so typically $i<4$).

In the Zee-Babu model \cite{Cheng:1980qt, Babu:1988ki, Babu:2002uu, McDonald:2003zj, AristizabalSierra:2006gb, Nebot:2007bc, Ohlsson:2009vk, 2014NuPhB.885..524S, Herrero-Garcia:2014hfa} neutrino masses are generated at two loops, where the new scalars cannot be very heavy or have very small Yukawa couplings, otherwise neutrino masses would be too small.

We follow the notation in \cite{Nebot:2007bc, Herrero-Garcia:2014hfa}, where a complete list of references is given. The Zee-Babu adds to the Standard Model two charged singlet scalar fields \begin{equation}
h^{\pm},\qquad k^{\pm\pm}\,,\end{equation}
with weak hypercharges $\pm 1$ and $\pm 2$ respectively.

The interesting Yukawa interactions are:
\begin{equation}
\mathcal{L}_{Y}=
\overline{\ell}\, Y e \phi +  \overline{\tilde{\ell}}f\ell h^{+}+\overline{e^{c}}g\, e\, k^{++}+
\mathrm{H.c.}\nonumber
\end{equation}
Due to Fermi statistics, $f_{ab}$ is an antisymmetric matrix in flavour space, while $g_{ab}$ is symmetric.

And the most general scalar potential has the form:
\begin{eqnarray}
V & = & m_{H}^{\prime2} H^{\dagger}H+ m_{h}^{\prime2}|h|^{2}+m_{k}^{\prime2}|k|^{2} \,+
\nonumber \\
&+&
\lambda_{H}(H^{\dagger}H)^{2}+\lambda_{h}|h|^{4}+\lambda_{k}|k|^{4} \,+
\nonumber \\
&+&\lambda_{hk}|h|^{2}|k|^{2} +  \lambda_{hH}|h|^{2}H^{\dagger}H+\lambda_{kH}|k|^{2}H^{\dagger}H\,+
\nonumber \\
&+&\left(\mu h^{2}k^{++}+\mathrm{H.c.}\right)\,, 
\label{eq:V}
\end{eqnarray}
being $H$ the $SU(2)$ doublet Higgs boson.

The potential has to be bounded from below, which requires that the quartic part should be positive for all values of the fields and for all scales. Then, if two of the fields $H,k$ or $h$ vanish one immediately finds
\begin{equation}
\label{eq:s1}
 \lambda_H > 0, \qquad \lambda_{h} > 0, \qquad  \lambda_{k} > 0  \, .
\end{equation}
Moreover the positivity of the potential whenever one of the scalar fields $H,h,k$ is zero implies  
\begin{equation}
\label{eq:s2}
\alpha,\beta,\gamma > -1\,,
\end{equation}
where we have defined $
\alpha = \lambda_{hH}/(2 \sqrt{ \lambda_H \lambda_{h}})$, 
$\beta = \lambda_{kH}/(2 \sqrt{ \lambda_H \lambda_{k}})$ and $\gamma = \lambda_{hk}/(2 \sqrt{ \lambda_h \lambda_{k}})$.

Eq.~\ref{eq:s2} constrains only negative mixed couplings, 
$\lambda_{xH}, \lambda_{hk}$ ($x=h,k$), since for positive ones  the potential is definite positive
and only the perturbativity limit, $\lambda_{xH},\lambda_{hk} \lesssim 4 \pi$ applies.
Finally, if at least two of the mixed couplings are negative, 
there is an extra constraint, which  can be written as:
\begin{equation}
\label{eq:s3} 
1-\alpha^2 -\beta^2 - \gamma^2 + 2 \alpha \beta \gamma > 0 
\qquad \vee \qquad \alpha+\beta+\gamma > -1\, . 
\end{equation} 

\begin{figure}[h]
	\centering
	\includegraphics[width=0.35\textwidth]{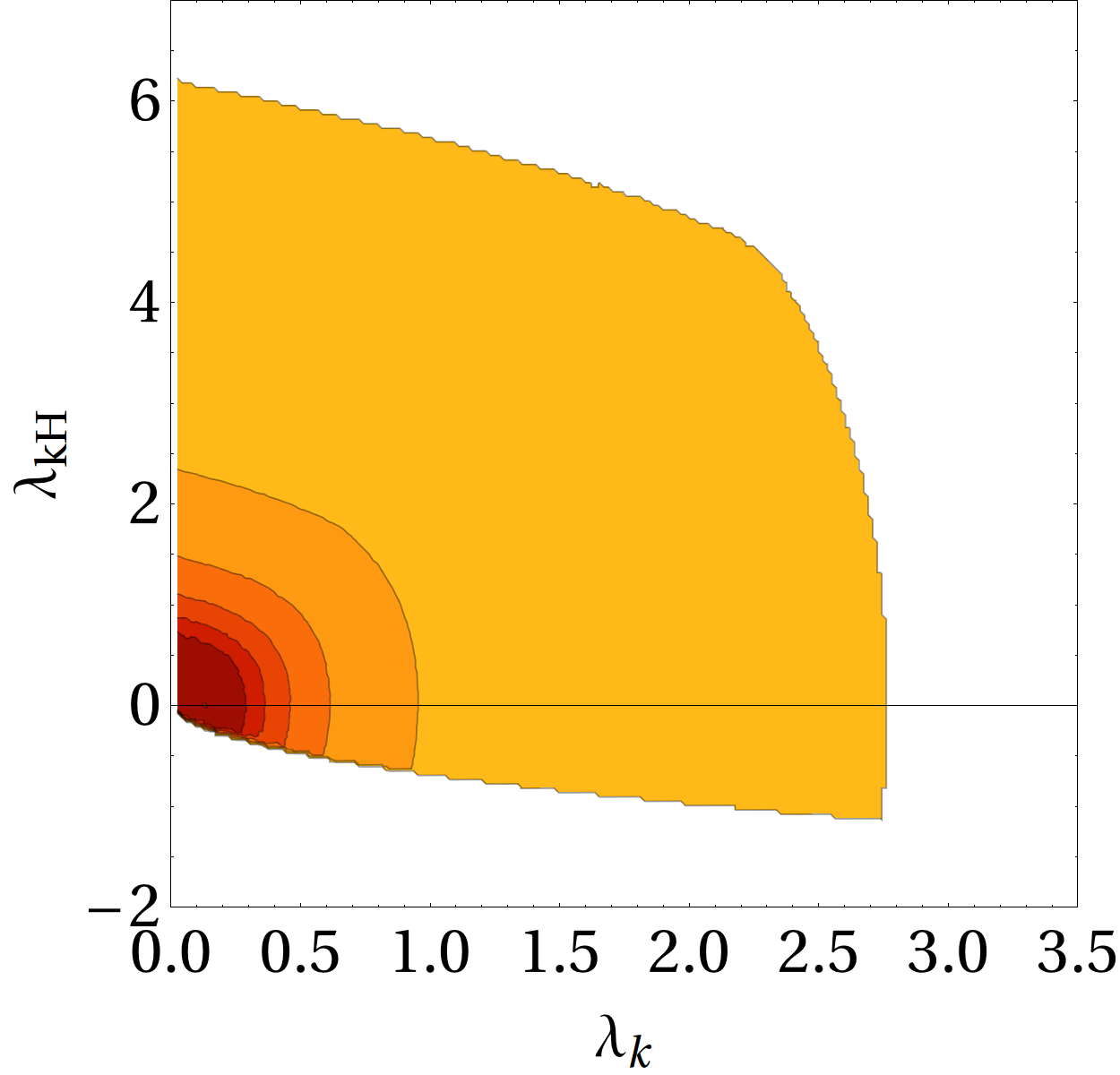}
		\caption{Allowed regions in $\lambda_{kH}$ vs $\lambda_k$, taken at the $m_Z$ scale, if perturbativity/stability is required to be valid up to $10^3,10^6,10^9,10^{12},10^{15},10^{18}$ GeV (from light to dark colours).} \label{fig:perturbativity}
\end{figure}

For a given set of parameters defined at the electroweak scale, 
and satisfying the stability conditions discussed above, 
one can calculate the running couplings numerically by using one-loop RGEs.
The new scalar couplings $\lambda_{hH},\lambda_{kH}$ 
always contribute positively to the running of the Higgs quartic coupling 
$\lambda_H$, compensating for the large and negative contribution of the top quark Yukawa coupling. 
Therefore, the vacuum stability problem can be alleviated in the ZB model. We show in figure \ref{fig:perturbativity} the allowed regions in the plane $\lambda_{kH}$ vs $\lambda_k$, at the EW scale, if perturbativity/stability is required to be valid up to a certain scale.

To see how neutrino masses can be generated in this model, it is important to remark that lepton number violation requires the simultaneous
presence of the four couplings $Y$, $f$, $g$ and $\mu$, because
if any of them vanishes one can always assign quantum numbers in such
a way that there is a global $U(1)$ symmetry.

\begin{figure}[h]
\begin{centering}
\includegraphics[width=0.9\columnwidth]{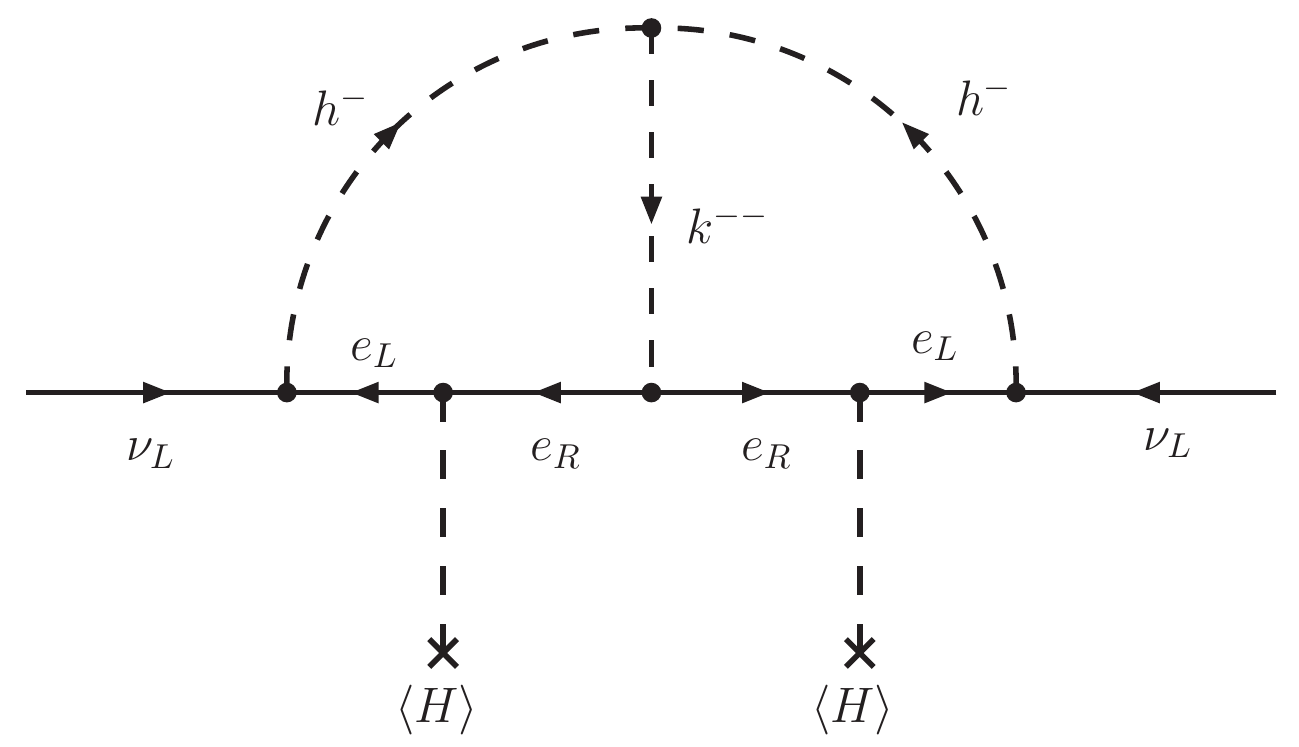}\par\end{centering}
\caption{Diagram contributing to Majorana neutrino masses at two loops.}
\label{fig:babumass}
\end{figure}

The neutrino masses (see fig.~\ref{fig:babumass}) can be written as:

 \begin{equation}
\mathcal{M}_{\nu}=\frac{v^{2}\mu}{48\pi^{2}M^{2}}\tilde{I}\, f\, Y\, g^{\dagger}Y^{T}f^{T}\,.\label{eq:MnuYukawas}
\end{equation}
where $\tilde{I}$ is an integral (see \cite{McDonald:2003zj,Nebot:2007bc, Herrero-Garcia:2014hfa} for details.).

A very important point is that since $f$ is a $3\times 3$ antisymmetric
matrix, $\det f=0$ (for $3$ generations), and therefore $\det\mathcal{M}_{\nu}=0$. Thus,
at least one of the neutrinos is exactly massless at this order.

\section{Constraints on the parameters of the model}

In principle, the scale of the new mass parameters of the ZB model 
($m_{h}, m_{k}$ and $\mu$) is arbitrary. However,  
 from the experimental point of view it is interesting to consider 
new scalars light enough to be produced in the LHC. Also 
theoretical arguments suggest that the scalar masses should be relatively 
light (few TeV), to avoid unnaturally large one-loop corrections to the Higgs mass \cite{Aad:2012tfa,Chatrchyan:2012ufa}
which would introduce a hierarchy problem.
Therefore, in this paper we will focus on the masses of the new scalars, 
$m_{h}, m_{k}$, below 2 TeV.

\begin{figure}[h]
\begin{centering}
\includegraphics[width=0.65\columnwidth]{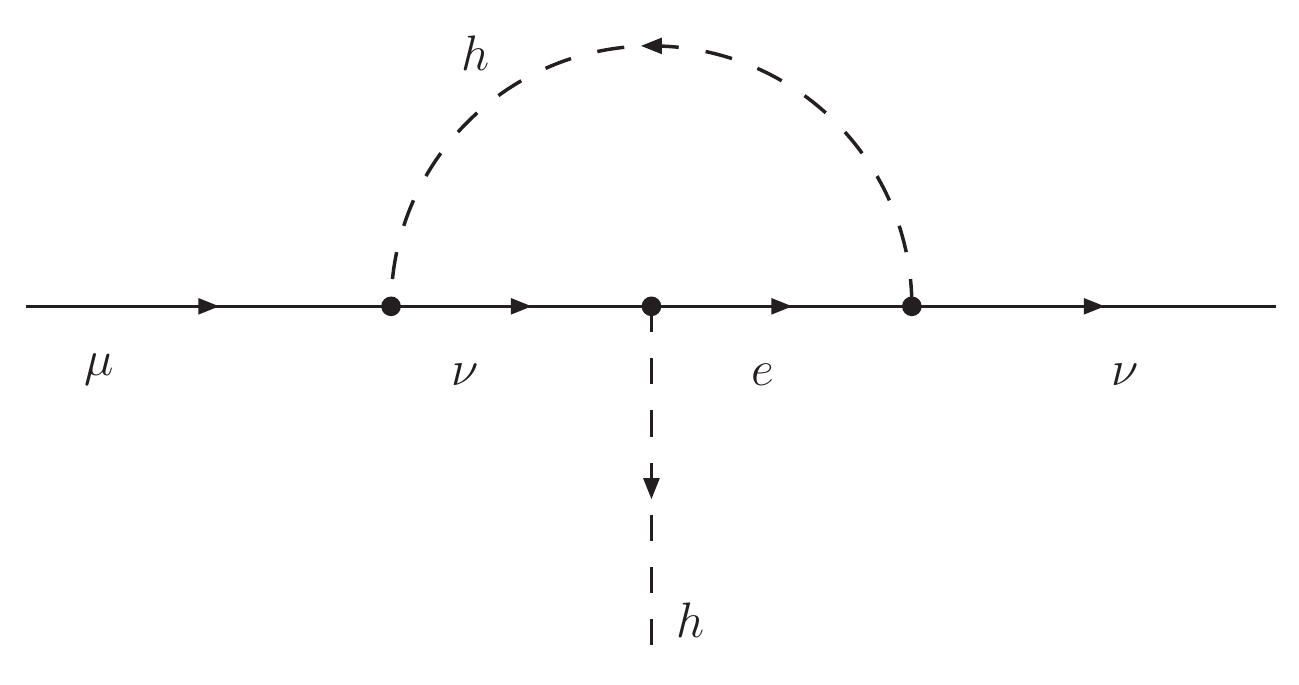}\
\includegraphics[width=0.65\columnwidth]{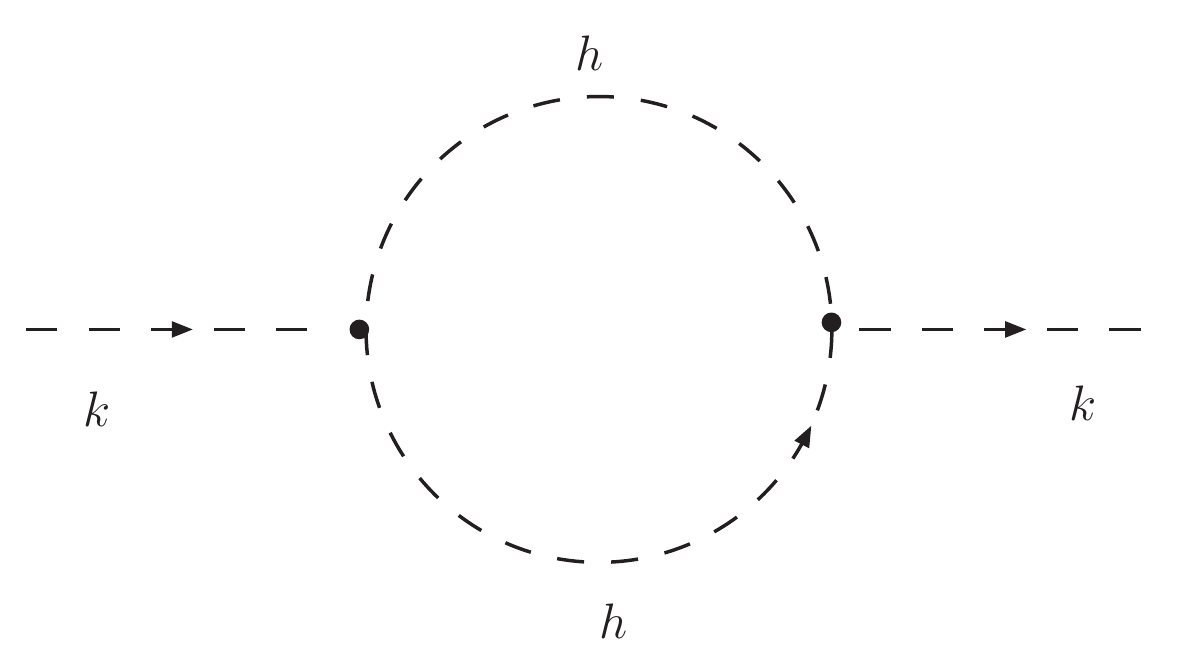}
\par\end{centering}
\caption{a) Top: one-loop corrections to the $h$ Yukawa couplings, $f$ (similarly for the $k$ ones, $g$). b) Bottom: 
contribution of $h$ to the self-energy of $k$. This constraints the $\mu$ parameter.}
  \label{selfk}
\end{figure}

Since one-loop corrections to Yukawa couplings (see figure \ref{selfk} (top)) are order
\begin{equation}
\delta f\sim\frac{f^{3}}{(4\pi)^{2}}\,,\qquad\delta g\sim\frac{g^{3}}{(4\pi)^{2}}\,
\end{equation}
one expects from perturbativity $f,g \ll 4\pi$.

\begin{figure}[h]
	\centering
	\includegraphics[width=0.23\textwidth]{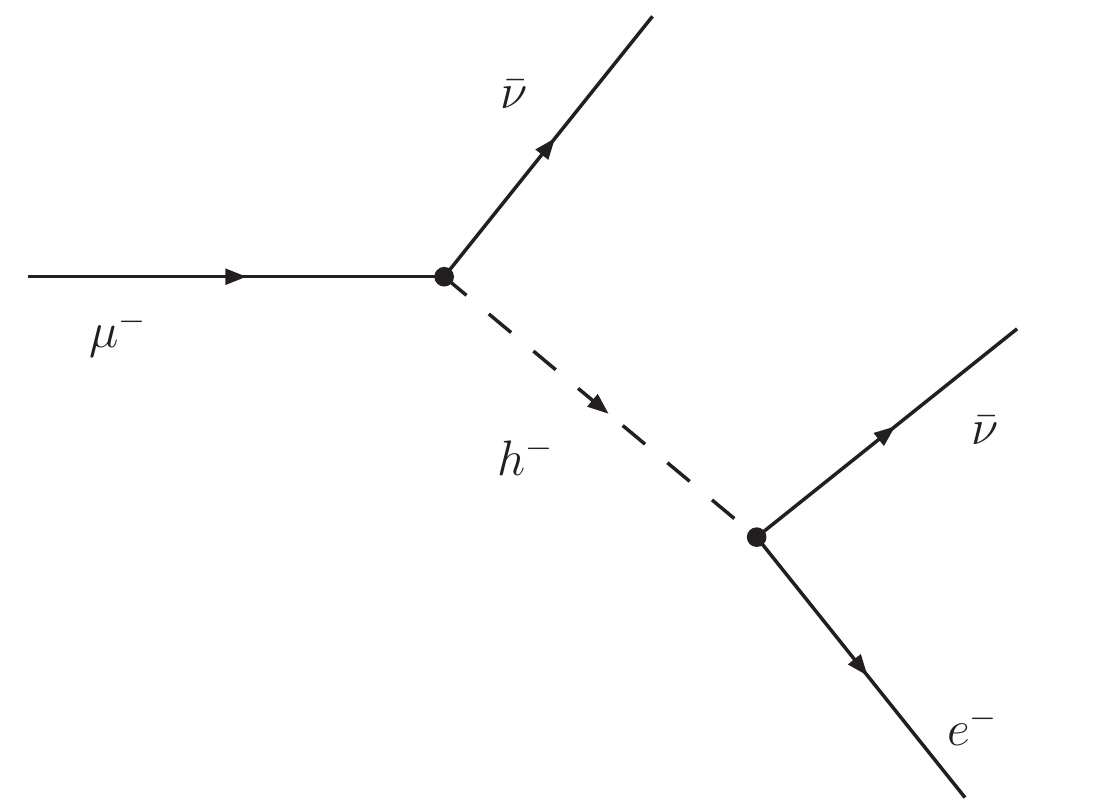}\\
	\includegraphics[width=0.23\textwidth]{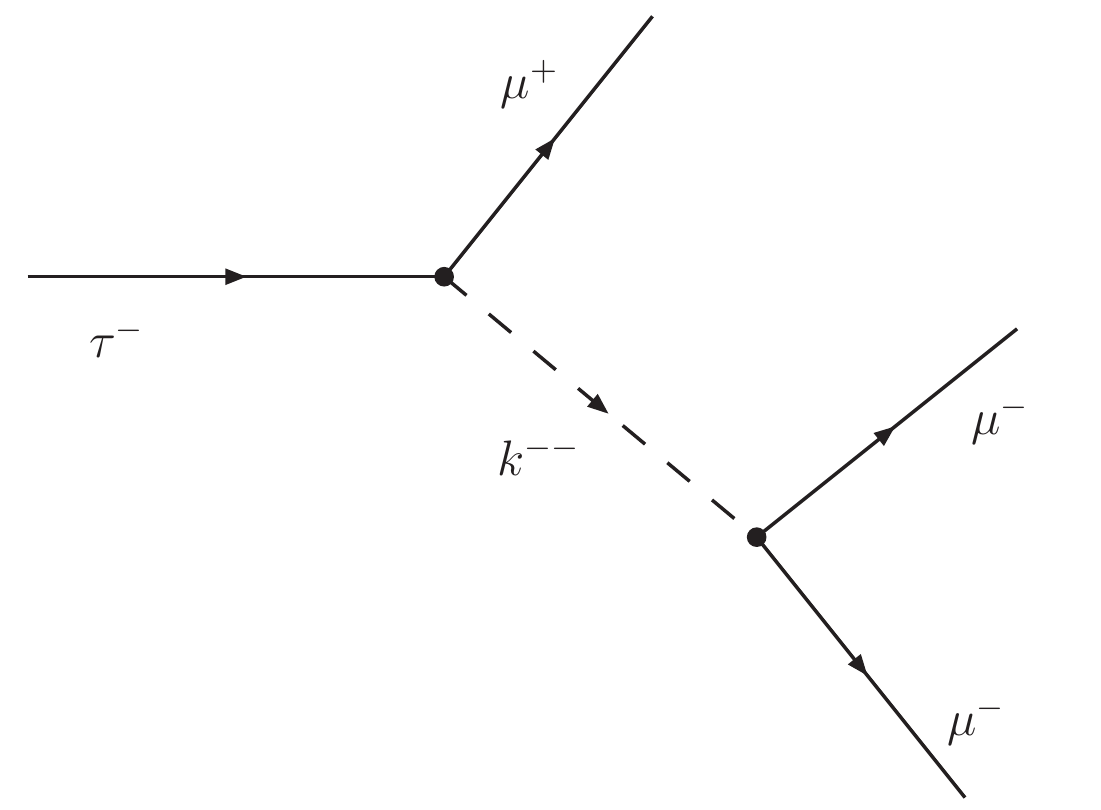}\\
	\includegraphics[width=0.28\textwidth]{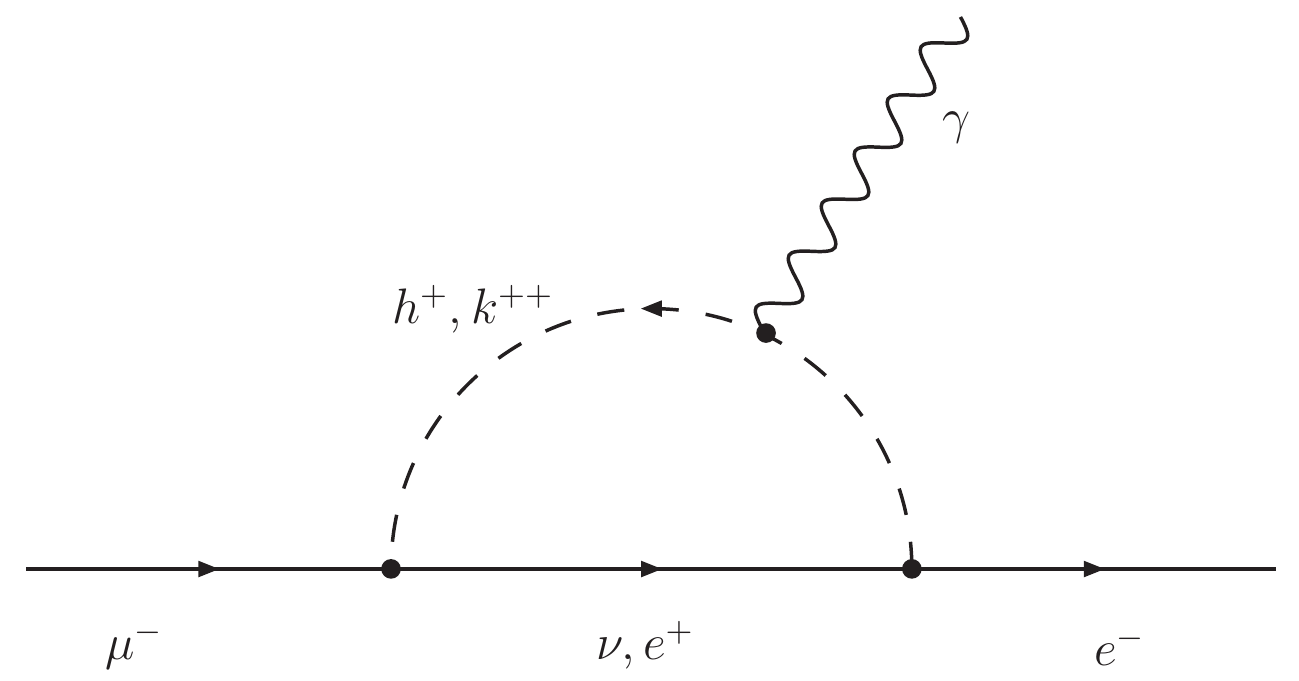}
	\caption{a) Top: tree-level lepton flavour violating decays mediated by the singly-charged $h$. The same type of diagram gives rise to non-standard interactions \cite{Ohlsson:2009vk, Herrero-Garcia:2014hfa}. b) Middle: tree-level lepton flavour violating decays ($\ell^-_a \rightarrow \ell^+_b  \ell^-_c  \ell^-_d$) mediated by the doubly-charged $k$. c) Bottom: one-loop level lepton flavour violating decays like $\ell^-_a \rightarrow  \ell^-_b \gamma$.} \label{munue}
\end{figure}

The trilinear coupling among charged scalars $\mu$, on the other hand, is different, for it has dimensions of mass and it is insensitive to high energy perturbative unitarity constraints. However, it induces radiative corrections (see figure \ref{selfk} (bottom)) to the masses of the charged scalars of order 
\begin{equation}
\delta m_{k}^{2},\delta m_{h}^{2}\sim\frac{\mu^{2}}{(4\pi)^{2}}\,.
\end{equation}
Requiring that the corrections in absolute value are much smaller than the masses (naturality)
we can derive a naive upper bound for this parameter. Given that the neutrino masses depend linearly on the parameter $\mu$, the viability of the model is  very sensitive to the upper limit allowed for $\mu$. 
Thus we choose to implement such limit in terms of a parameter $\kappa$, 
\begin{equation} \label{mu}
\mu< \kappa \, {\rm min} (m_{h},m_k)  \,,
\end{equation}
and discuss our results for different values: $\kappa=1,5,4\pi$.

Several low-energy processes bound the couplings of the model \cite{Beringer:1900zz,Adam:2013mnn}, like those plotted in figure \ref{munue}:
\begin{itemize}

\item $\ell_a \rightarrow \ell_b \nu \bar{\nu}$ processes bound the $f_{ab}$ couplings, see fig. \ref{munue}a (top). For instance, from universality, $\sum_q |V_{uq}|^{2}=0.9999\pm0.0006$, we get:
\be 
|f_{e\mu}|^{2}<0.007\,\left(\frac{m_{h}}{\mathrm{TeV}} \right)^{2}.
\ee
The same type of diagram gives rise to non-standard interactions, which however are at most at the level of $\sim 10^{-4}$, and are therefore difficult to test \cite{Ohlsson:2009vk, Herrero-Garcia:2014hfa}.

\item $\ell_a^- \rightarrow \ell_b^+ \ell_c^- \ell_d^-$ processes bound the $g_{ab}$ couplings, see fig. \ref{munue}b (middle). For instance, 
$BR (\mu^{-}\rightarrow e^{+}e^{-}e^{-})<1.0\times10^{-12}$ \cite{Beringer:1900zz} gives:
\be
|g_{e\mu}g_{ee}^{*}|<2.3\times10^{-5}\,\left(\frac{m_{k}}{\mathrm{TeV}} \right)^{2}.
\ee

\item $\ell_a^- \rightarrow \ell_b^- \gamma$ processes bound the $f_{ab}$ and the $g_{ab}$ couplings, see fig. \ref{munue}c (bottom). For instance, $BR(\mu\rightarrow e\gamma)<5.7\times10^{-13}$ \cite{Adam:2013mnn} implies:
\end{itemize}
\be
\frac{|f_{e\tau}^{*}f_{\mu\tau}|^{2}}{(m_{h}/\mathrm{TeV})^{4}}+\frac{16|g_{ee}^{*}g_{e\mu}+g_{e\mu}^{*}g_{\mu\mu}+g_{e\tau}^{*}g_{\mu\tau}|^{2}}{(m_{k}/\mathrm{TeV})^{4}}< \frac{1.6}{10^{6}}
\ee

Similar constraints exist for other combinations of couplings, see reference \cite{Herrero-Garcia:2014hfa} for the complete list of processes and bounds on the couplings and masses.

 There are also predictions for lepton number violating processes, in particular from $0\nu \beta\beta$, which has just the typical light neutrino contribution (with $m_{\rm lightest}\sim0$):
  \begin{itemize}
\item  In Normal Hierarchy, 
\begin{equation}
(\mathcal{M}_{\nu}^{NH})_{ee}=\sqrt{\Delta_{S}}c_{13}^{2}s_{12}^{2}e^{i\phi}+\sqrt{\Delta_{A}}s_{13}^{2}
\,.\nonumber
\end{equation}
One obtains $0.001  \lesssim \,\mathrm{eV} |(\mathcal{M}_{\nu}^{NH})_{ee}| \lesssim 0.004\,\mathrm{eV}$, outside reach of planned experiments.

\item In Inverted Hierarchy,  
  \end{itemize}
\begin{equation}
(\mathcal{M}_{\nu}^{IH})_{ee}=\sqrt{\Delta_{A}+\Delta_{S}}c_{13}^{2}s_{12}^{2}e^{i\phi}+\sqrt{\Delta_{A}}c_{13}^{2}c_{12}^{2}\,.
\nonumber\end{equation}
One gets $0.01\,\mathrm{eV}  \lesssim |(\mathcal{M}_{\nu}^{NH})_{ee}| \lesssim 0.05\,\mathrm{eV}$, which is in the observable range of planned experiments.

Analytically, once all these constraints have been taken into account, we can plug them into eq. \ref{eq:MnuYukawas} and do a naive estimate of the allowed scalar masses:
 \begin{equation} 
\frac{m_{33}}{0.05 \, \mathrm{eV}} \simeq  500   |g_{\mu\mu}| |f_{\mu\tau}|^2  \frac{\mu}{M}
\frac{\mathrm{TeV}}{M} ,
\end{equation}
which implies:
\be
 m_{h,k} \gtrsim \frac{1 \,(3)\, \mathrm{TeV}}{\sqrt{\kappa}} \, {\rm NH\,(IH)}.
\end{equation}

As will be shown, cancellations in $m_{\alpha\beta}$ may occur depending on the values of the phases, and these lead to much lower scalar masses allowed, especially for IH.

\section{Numerical scan}

To check the analytical estimates of the model, we have performed a MCMC. There are $17$ parameters:
\begin{itemize}
\item $9$ moduli: $3$ from $f$ 
 and $6$ from $g$. 
\item $5$ phases: $3$ from $g$ and $2$ from $f$.
 \item the real and positive parameter $\mu$, and $m_h, m_k$, the physical masses of the scalars.
 \end{itemize}

We fix 2 $f's$ and 3 $g's$ in terms of the neutrino mixing angles and masses, which we vary in their 1$\sigma$ range, and we take as independent phases those of the $g_{e\alpha}$ couplings plus the Majorana and the Dirac phases of the neutrino mixing matrix.
The following table summarizes the allowed range of variation of the parameters: 

 \begin{table}[ht]
\begin{centering}\begin{tabular}{||c|c|c||}
\hline 
Parameter &
Allowed range \\
\hline
\hline 
$\Delta_{S}$ &
\, $(7.50\pm0.19)\,10^{-5}\,\mathrm{eV^{2}}$\, \\
\hline 
$\Delta_{A}$ &
$(2.45 \pm 0.07)\,10^{-3} \mathrm{eV^{2}}$\\ 
\hline 
$\sin^{2}\theta_{12}$ &
$0.30\pm0.13$\\
\hline 
$\sin^{2}\theta_{23}$ &
$(0.42\cup 0.60)\pm 0.04$\\ 
\hline 
$\sin^{2}\theta_{13}$ &
$0.023\pm0.002$\\ 
\hline
$\delta, \phi$ &
$[0, 2\pi]$\\ 
\hline 
\, $\arg(g_{ee}),\arg(g_{e\mu}),\arg(g_{e\tau})$ \, &
$[0, 2\pi]$ \\ 
\hline 
$f_{\mu\tau},|g_{ee}|,|g_{e\mu}|,|g_{e\tau}|$ &
$[10^{-7}, 5]$\\ 
\hline 
$m_{h}$ &
$[100,  2 \times10^{3}]\,\mathrm{GeV}$\\
\hline 
$m_k$ &
$[200,  2 \times10^{3}]\,\mathrm{GeV}$\\
\hline 
$\mu$ &
$[1, 2\kappa\times10^{3}]\,\mathrm{GeV}$\\ 
\hline 
\end{tabular}\par\end{centering}
\end{table}

 There are many parameters and few observations, being most of them bounds. We have implemented these bounds in the numerical analysis with a constant and a Gaussian part, which avoids imposing stepwise bounds or half-Gaussian with best value at zero that penalize deviating from null when this might not be supported.

For an experimental bound $B^{\mathcal O}_{[\mathrm{90\,\%CL}]}$ at 90\% CL (1.64$\sigma$), the $\chi^2$ contribution of the observable $\mathcal O_{\rm th}$ is
\[
\chi^2(\mathcal O_{\rm th})=\left\{\begin{array}{l}0,\  \mathcal O_{\rm th}<\frac{B^{\mathcal O}_{[\mathrm{90\,\%CL}]}}{1.64},\\ \left(\frac{1.64\mathcal O_{\rm th}}{B^{\mathcal O}_{[\mathrm{90\,\%CL}]}}-1\right)^2\left(\frac{1.64}{0.64}\right)^2,\ \mathcal O_{\rm th}\geq \frac{B^{\mathcal O}_{[\mathrm{90\,\%CL}]}}{1.64}.\end{array}\right.
\]

\begin{figure}[h] 
	\centering
	\includegraphics[width=0.35\textwidth]{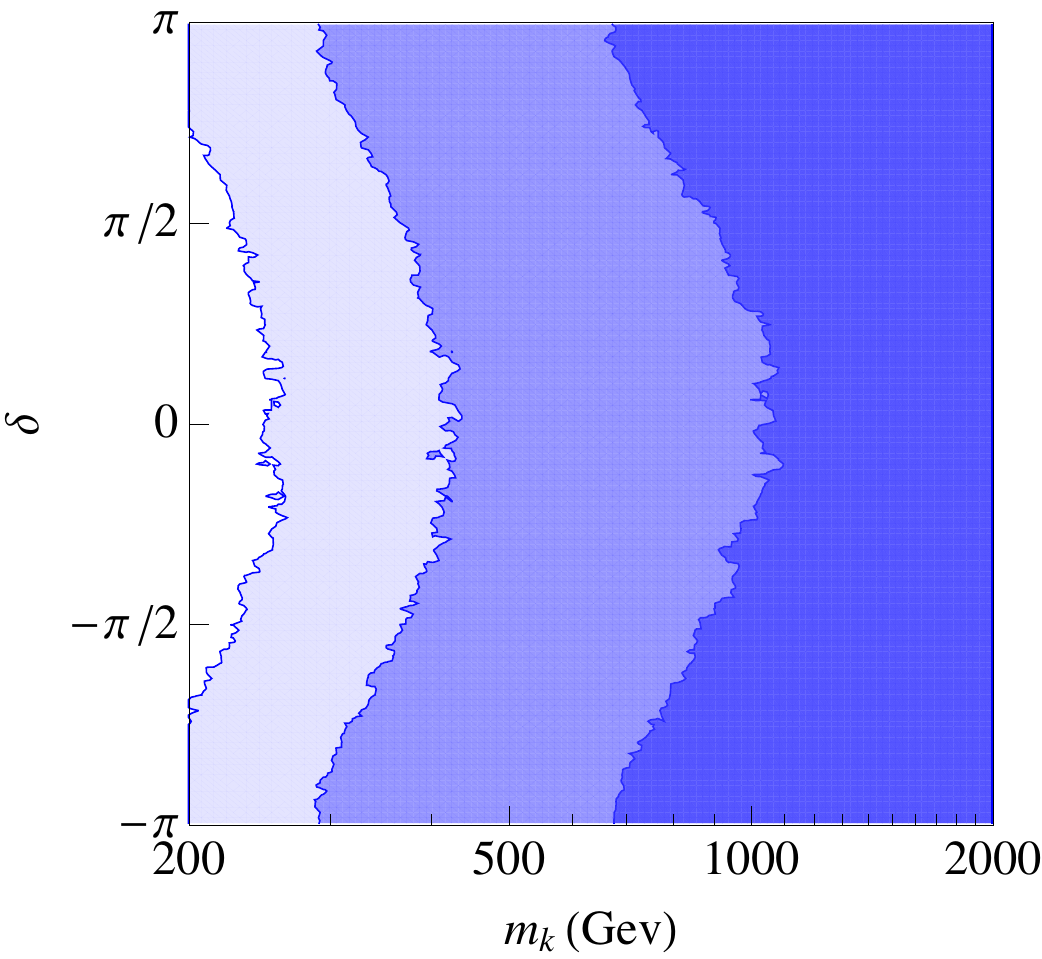}\\
	\includegraphics[width=0.35\textwidth]{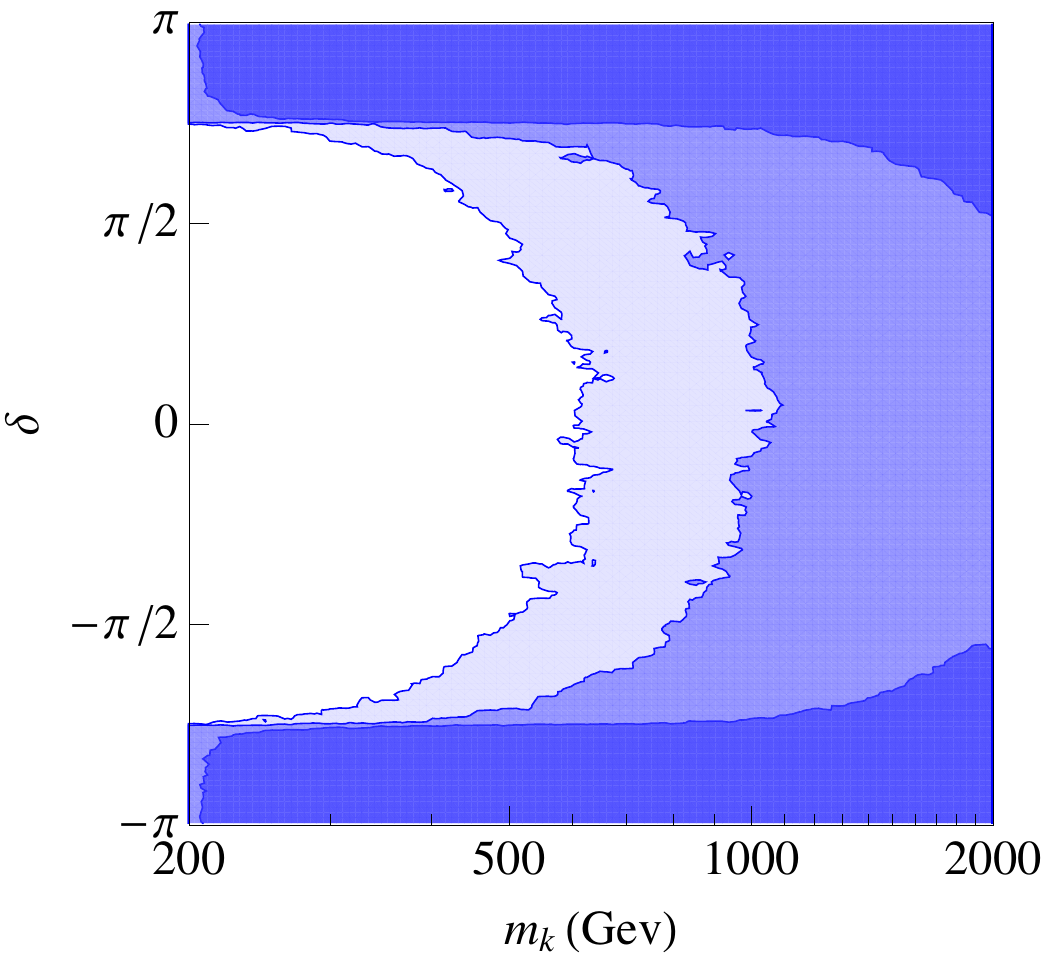}
	\caption{$\delta$ vs $m_k$ in NH (top) and IH (bottom).} 
	\label{delta-mk}
\end{figure}

In the  plots we show the regions with the total $\Delta \chi^2 \leq 6$, which corresponds to 95\% confidence levels with two variables. In figure \ref{delta-mk} we show the plane $\delta$ versus $m_k$ in NH (top) and IH (bottom), for different values of $\kappa$: $1,5,4\pi$ from dark to light blue.

Similar plots exist for $m_h$, and also versus the Majorana phase $\phi$. As can be seen, the scalar masses are accessible at LHC-14 depending on the values of the phases. They are mainly produced via Drell-Yan processes. Their dominant decay channels, however, will depend on the neutrino mass hierarchy.

\begin{figure}[h]
	\centering
	\includegraphics[width=0.33\textwidth]{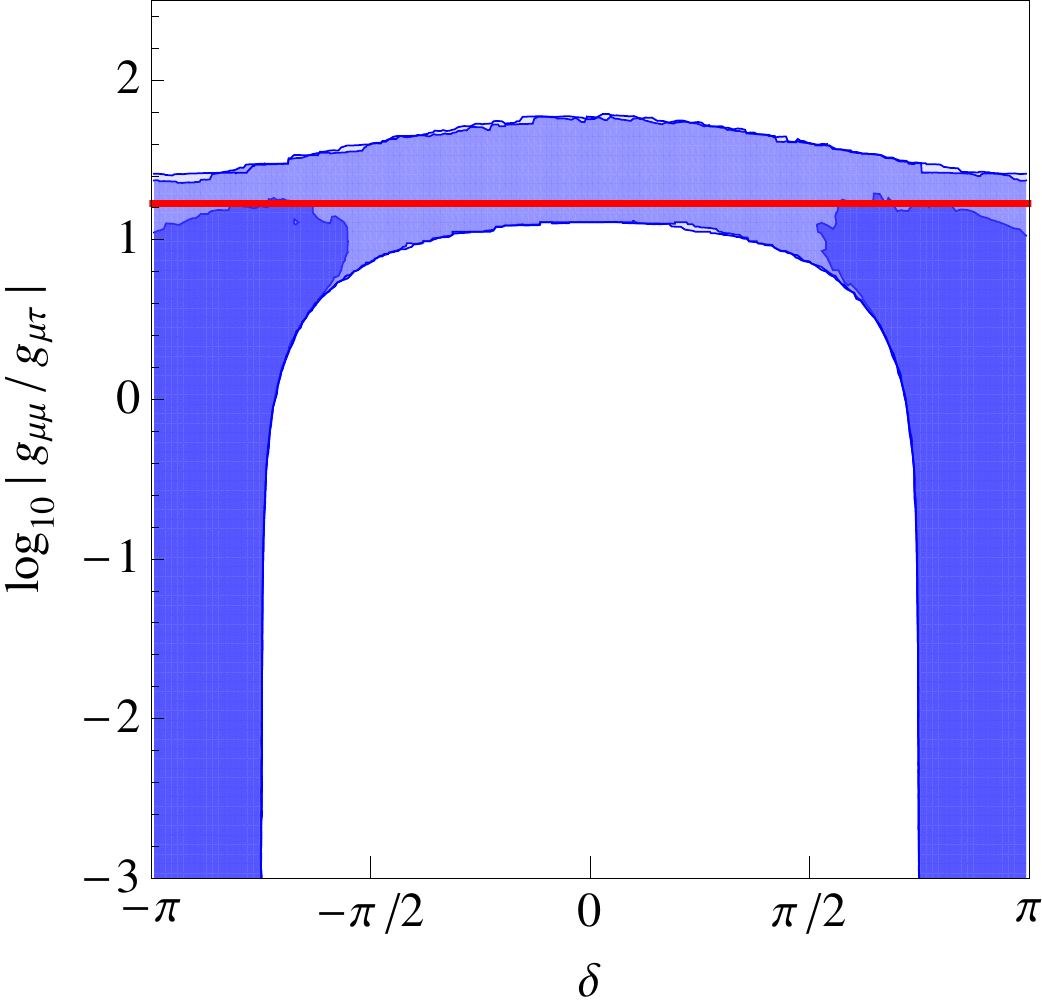}\\
	\includegraphics[width=0.33\textwidth]{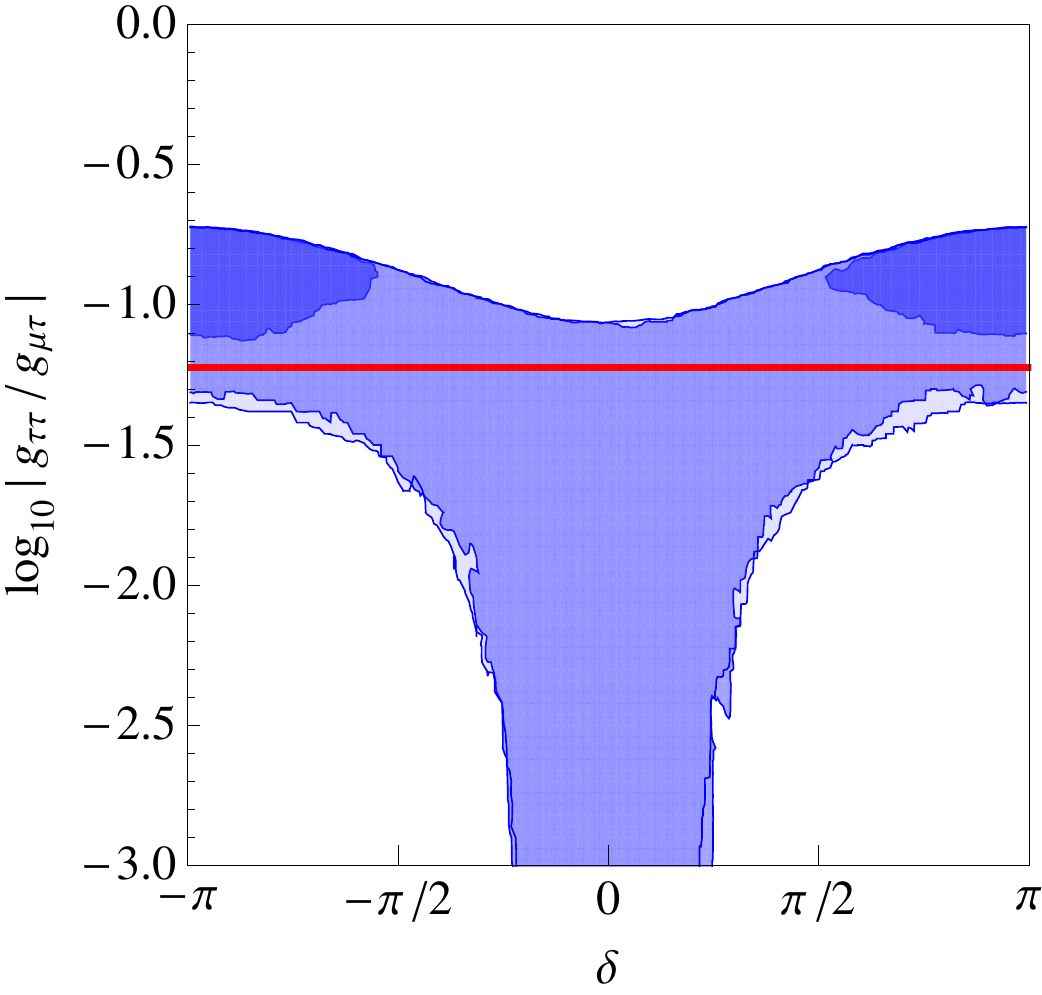}
	\caption{$\log|g_{\mu\mu}/g_{\mu\tau}|$ (top)and $\log|g_{\tau\tau}/g_{\mu\tau}|$ (bottom) vs $\delta$ for IH. The horizontal red lines represent the naive approximation.}
	 \label{gmmIH}
	 \end{figure}

In the Zee-Babu model, from eq. \ref{eq:MnuYukawas}, we know that 
neglecting $m_e \ll m_\mu,m_\tau$:
 \begin{equation} \nonumber \label{m22}
m_{22} \propto f_{\mu\tau}g_{\tau\tau} m_\tau^2,
\end{equation}
   \begin{equation} \nonumber  \label{m23}
m_{23} \propto f_{\mu\tau}g_{\mu\tau} m_\mu m_\tau,
\end{equation}
\begin{equation} \nonumber  \label{m33}
 m_{33} \propto f_{\mu\tau} g_{\mu\mu} m_\mu^2,
 \end{equation}
and as the atmospheric angle is large, $m_{22}\sim m_{23}\sim m_{33}$ implies a naive scaling $g_{\mu\mu} / g_{\mu\tau} \sim g_{\mu\tau} / g_{\tau\tau}Ê\sim  m_{\tau}/m_{\mu}$. We have checked that this is indeed fulfilled for NH, while for IH it is not, as can be seen in figure \ref{gmmIH}, in which we plot, for IH, $\log|g_{\mu\mu}/g_{\mu\tau}|$ and $\log|g_{\tau\tau}/g_{\mu\tau}|$ vs $\delta$, and the naive scaling as red lines.

\begin{figure}[h]
	\centering
	\includegraphics[width=0.4\textwidth]{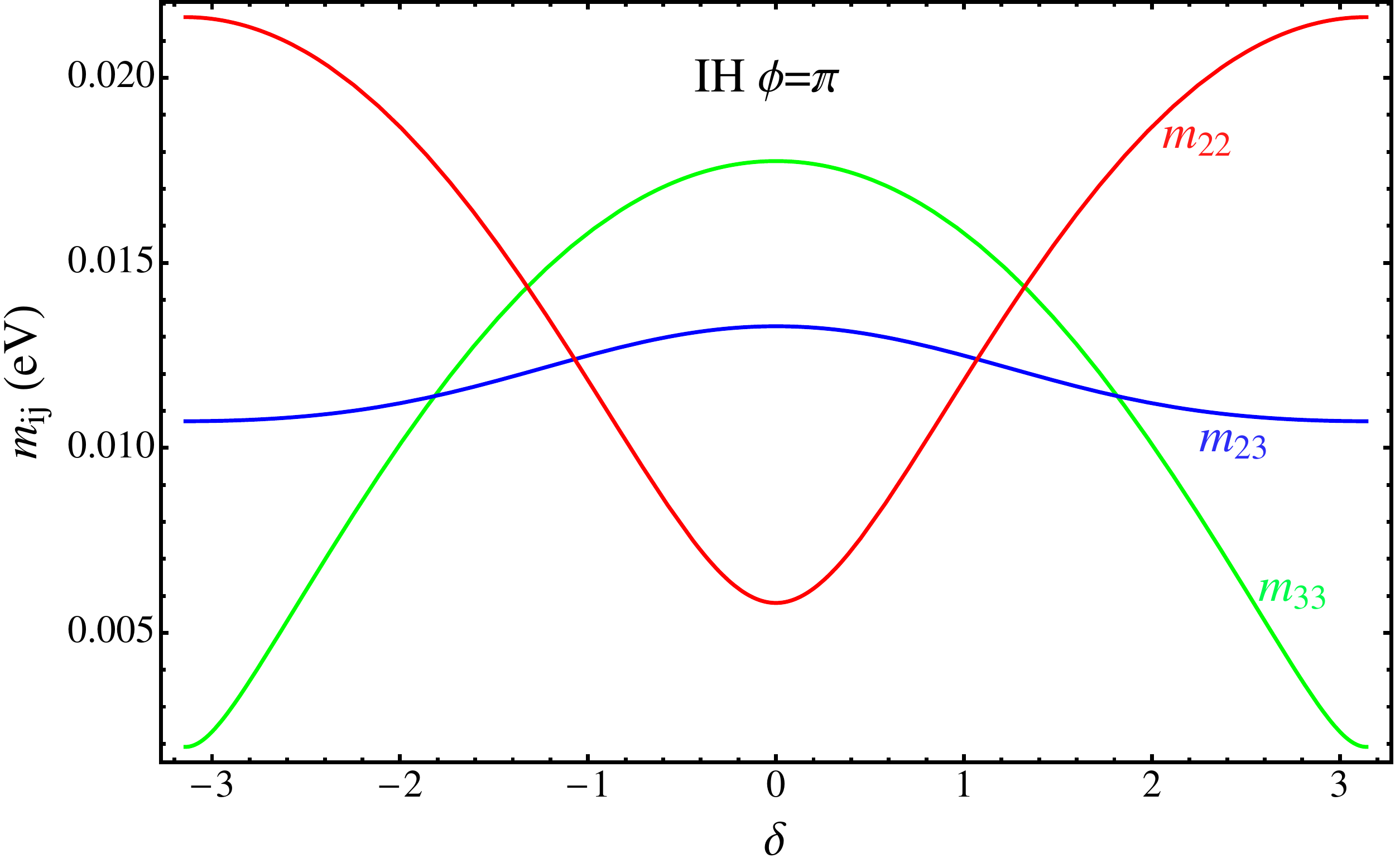}~~
	\caption{$m_{22}, m_{23}, m_{33}$ versus $\delta$ (IH), for fixed Majorana phase $\phi=\pi$.}
	 \label{pZB}
\end{figure}

This can be understood from eqs. \ref{m22},  \ref{m23} and \ref{m33} and figure \ref{pZB}, which shows how $m_{33}$ is the smallest for $\delta\sim \pi$, and $m_{22}$ for $\delta\sim 0$.

 \begin{figure}[h]
\centering 
	\includegraphics[width=0.35\textwidth]{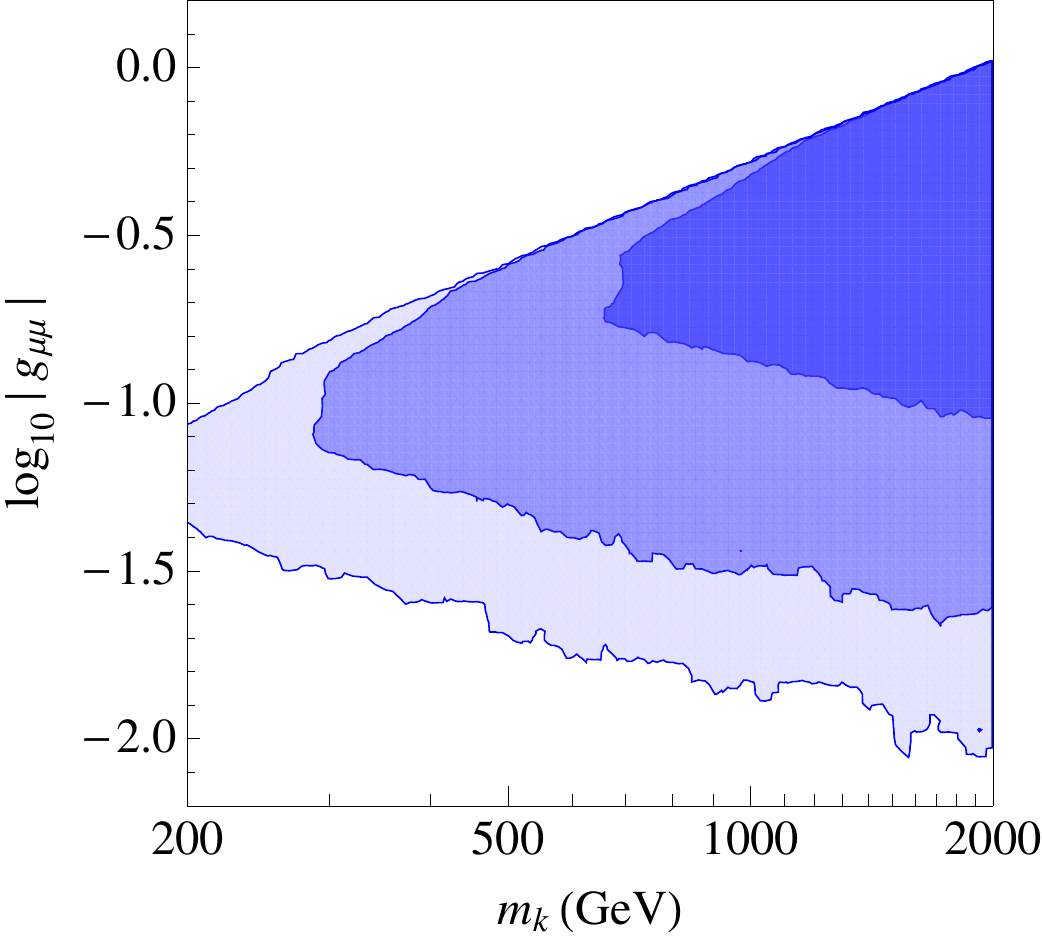}\\
	\includegraphics[width=0.35\textwidth]{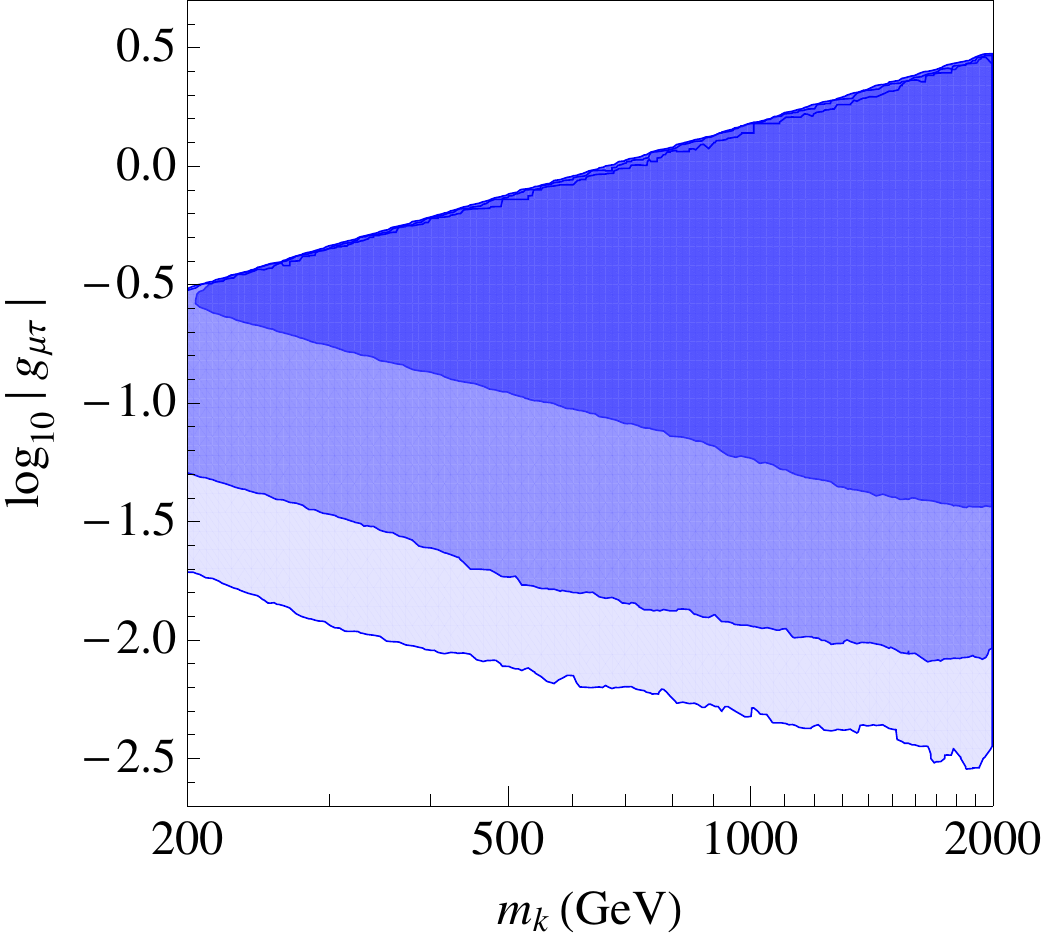}
	\caption{$\log|g_{\mu\mu}|$ vs $m_k$ for NH (top) and $\log|g_{\mu\tau}|$ vs $m_k$ for IH (bottom).}
	\label{gmk}
\end{figure}

In figure  \ref{gmk} we show the largest of the couplings $g_{\alpha\beta}$ versus $m_k$:  $g_{\mu\mu}$ for NH (top) and $g_{\mu\tau}$ for IH (bottom).
Together with $k \rightarrow e e $, these are the most promising channels for the LHC to detect the doubly-charged scalar. 

\begin{figure}[h]
	\centering
	\includegraphics[width=0.32\textwidth]{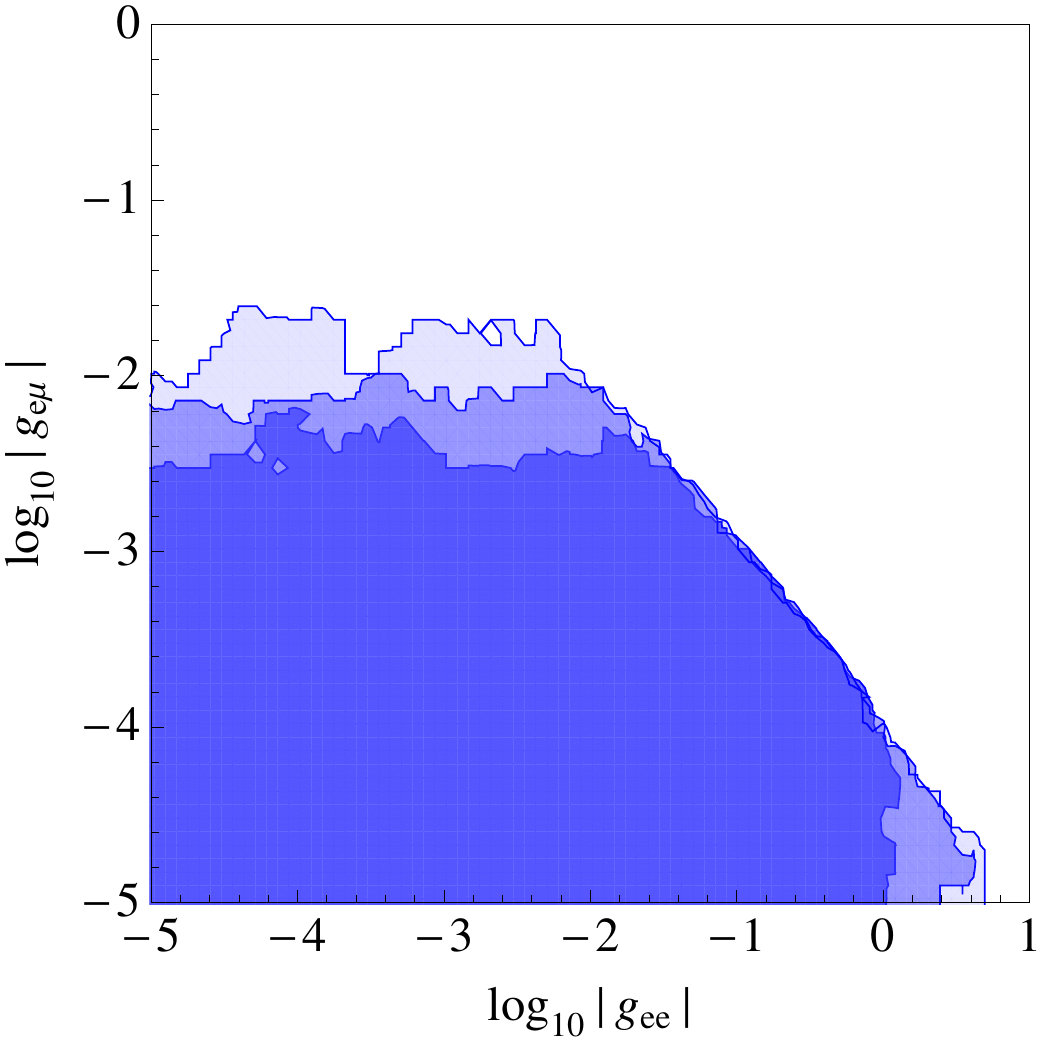}\\
	\includegraphics[width=0.32\textwidth]{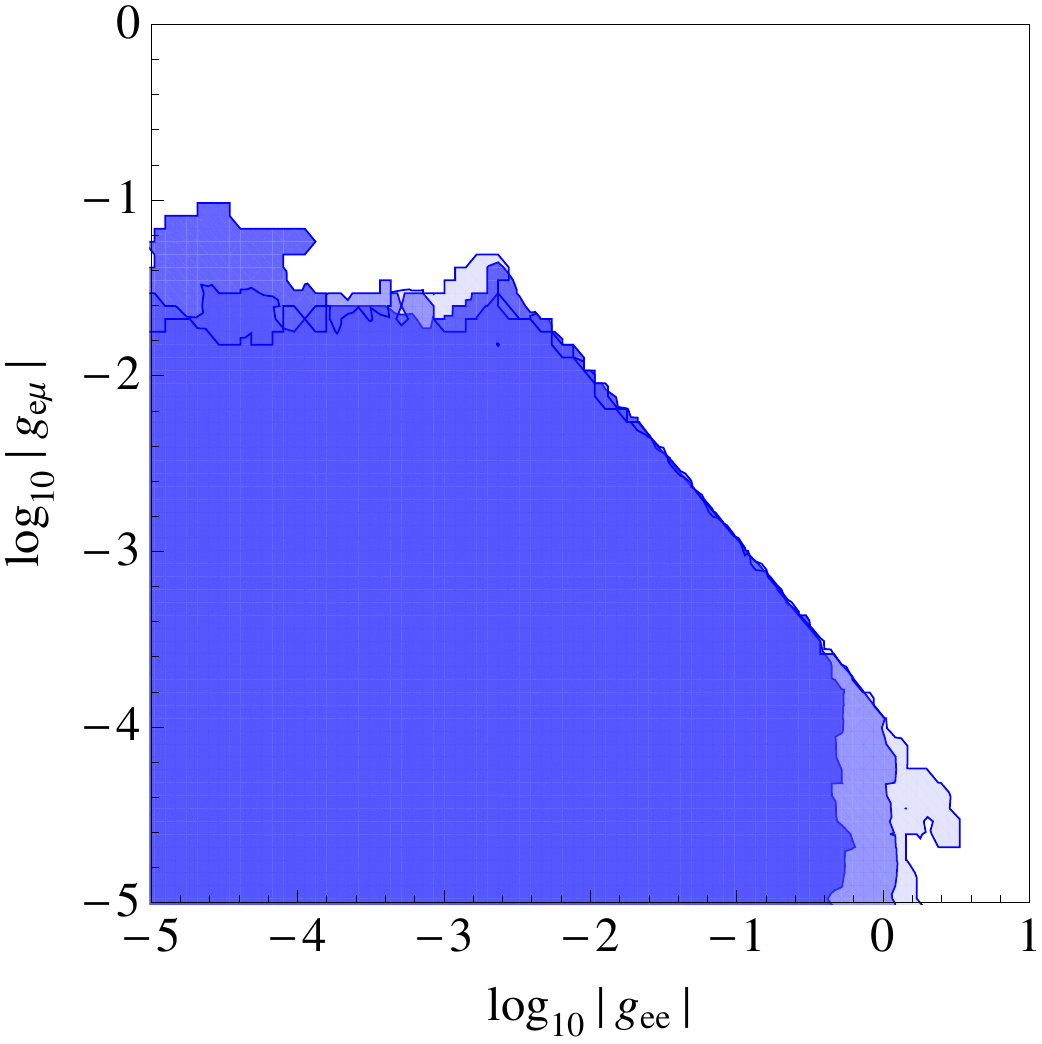}
	\caption{$\log|g_{e\mu}|$ vs $\log|g_{ee}|$ for NH (top) and IH (bottom).} 
	\label{fig:gem-gee}
\end{figure}

 In both hierarchies BR($k \rightarrow e\mu,e\tau,\tau\tau$) are negligible. For instance, the constraint on 
$|g_{ee} g_{e\mu}|$  from $\mu\rightarrow 3e$  implies that 
$|g_{ee} g_{e\mu}|<2.3\times10^{-5}\,(m_{k}/ \mathrm{TeV})^{2}$, as seen in fig.~\ref{fig:gem-gee}.

In IH one has to take into account that the channel $k \rightarrow h h$ is open, and this means that LHC-14 limits will not directly apply, as detecting the singly-charged $h$ is experimentally much harder.
  
 \begin{figure}[h]
\begin{centering}
\includegraphics[width=0.7\columnwidth]{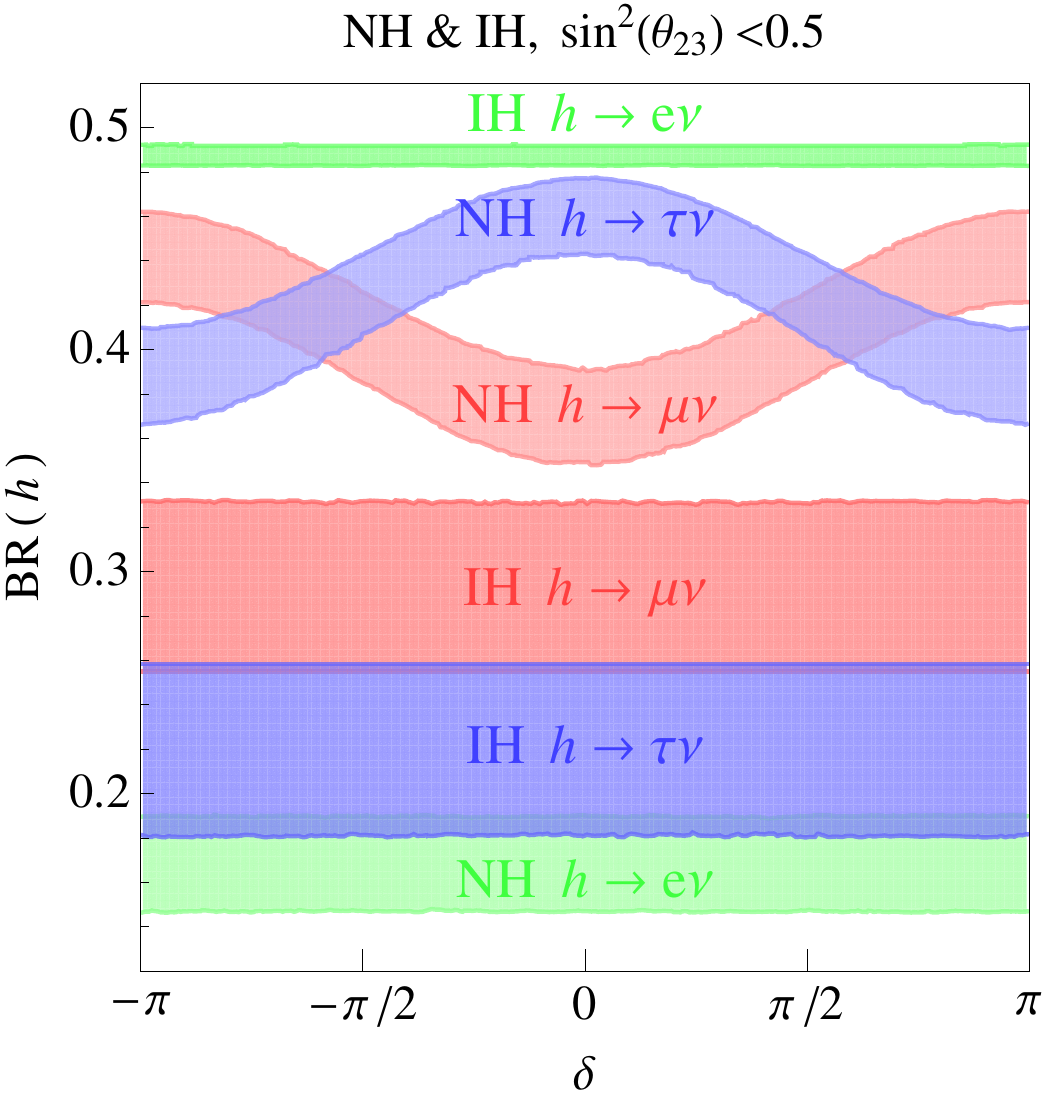}\\
\includegraphics[width=0.7\columnwidth]{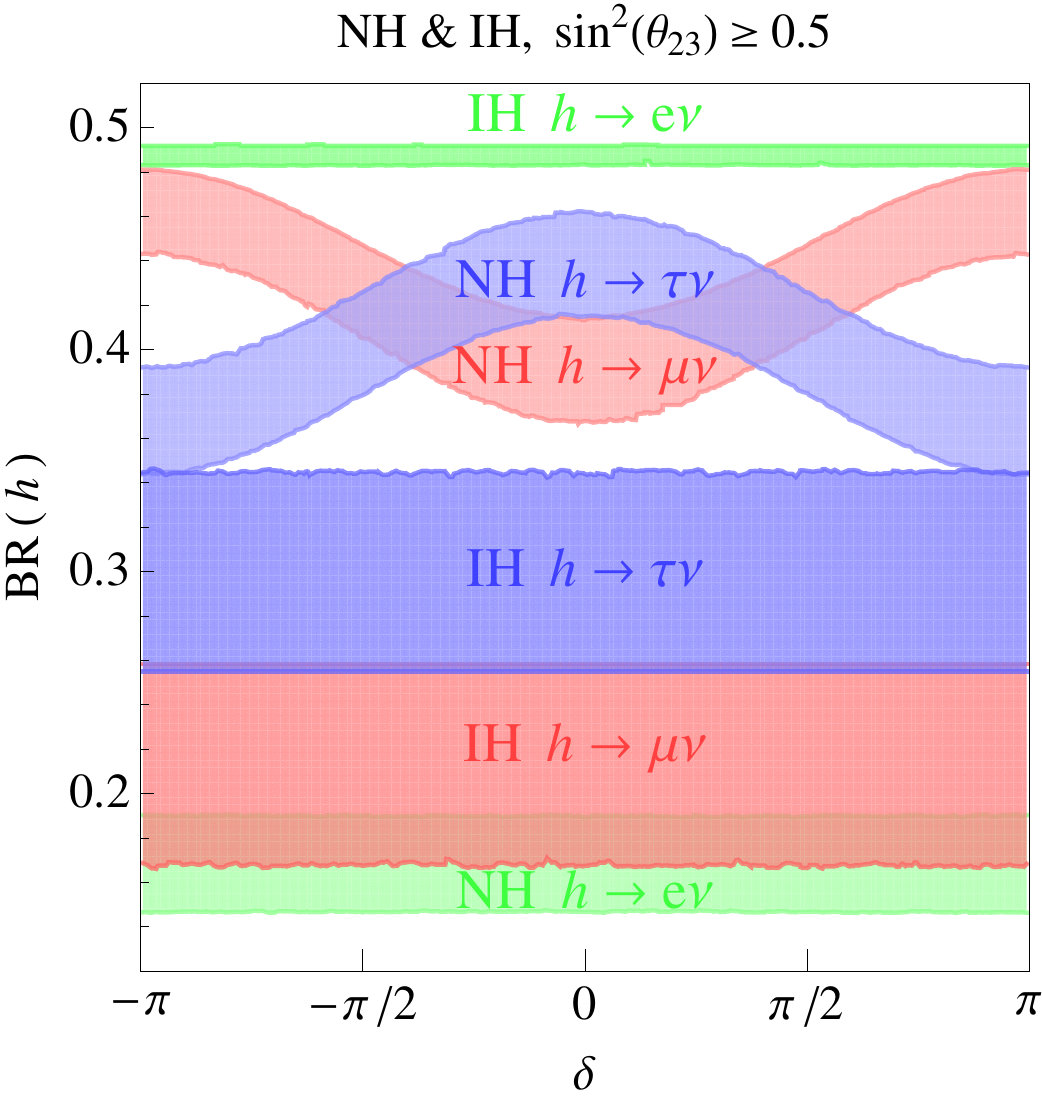}
\caption{Branching ratios of the charged singlet $h$ to $e\nu, \mu\nu,\tau\nu$ for $\theta_{23}$, $\theta_{23} < 45^\circ$ ($\theta_{23} > 45^\circ$) top (bottom). One can see the dependence on $\delta$ for the NH spectrum in the $\mu\nu$ and $\tau\nu$ channels. The bands are $95$\% C.L. regions.}
 \label{BRh}
\par\end{centering}
\end{figure}

We show in figure \ref{BRh} the branching ratios of $h$, BR($h\rightarrow e\nu, \mu\nu,\tau\nu$), for $\theta_{23} < 45^\circ$ (top) and $\theta_{23} > 45^\circ$ (bottom).
 It will be difficult to test, as we always have a large SM background, like $W\rightarrow e\nu$. If detected, the $e\nu$ channel is the best option to discriminate between hierarchies. Notice also that there is a mild dependence on $\delta$ in the $\mu\nu$ and $\tau\nu$ channels for NH.

\section{Conclusions}

We have studied the ZB model in the light of the second run of the LHC, taking into account the new available data from low-energy and neutrino experiments, and from direct searches. There are a number of important points and conclusions worth remarking:
 \begin{itemize}
  \item There are many LFV processes, like $\mu \rightarrow 3\,e $, $\mu \rightarrow e \gamma$ and $\mu - e$ conversion, $\tau$ decays... which are probing the model, but which by themselves cannot rule it out.
  \item One of the most important predictions of the model is that there is one neutrino with $m\sim 0$, so that the neutrino spectrum cannot be degenerate.
   \item In general,  the scalar masses are accessible to LHC-14 in both hierarchies, but input information from neutrino experiments (hierarchy, $\delta$, $\theta_{23}$ octant) is crucial to really pin-down the model. In fact, if the spectrum is inverted and $\delta$ is quite different from $\sim\pi$, the scalars will be outside LHC-14 reach.
 \item If any of the singlets is discovered, the model can be falsified using their decay modes and neutrino data (spectrum, $\delta$, $\theta_{23}$...). Complementarity of both direct and indirect searches is of uttermost importance.
  \end{itemize}

\section*{Acknowledgments}
This work has been partially supported by the  European Union FP7  ITN INVISIBLES (Marie Curie Actions, PITN- GA-2011- 289442), by the Spanish MINECO under grants  FPA2011-23897, FPA2011-29678, Consolider-Ingenio PAU (CSD2007-00060) and CPAN (CSD2007- 00042) and by Generalitat Valenciana grants PROMETEO/2009/116
and PROMETEO/2009/128.  M.N. is supported by a postdoctoral fellowship of project CERN/FP/123580/2011 at CFTP (PEst-OE/FIS/UI0777/2013), projects granted by \emph{Funda\c{c}\~ao para a Ci\`encia e a Tecnologia} (Portugal), and partially funded by POCTI (FEDER).




\nocite{*}
\bibliographystyle{elsarticle-num}
\bibliography{ZB-nuphbp1.bib}







\end{document}